\documentstyle[preprint,aps,epsf]{revtex}
\def\centerplot#1#2{
\begin{center}
\leavevmode
{\epsfxsize=#2\textwidth \epsfbox{#1}}
\end{center}
}
\begin{document}
\draft
\title{
Strong coupling theory of the interlayer tunneling model\\
for high temperature superconductors
}
\author{ Bo\v zidar Mitrovi\' c and Melissa Castle}
\address{
Department of Physics, Brock University, St. Catharines, Ontario,
Canada L2S 3A1
}
\maketitle
%
%
%
%
%
\begin{abstract}

The interlayer pair tunneling model of Anderson {\em et al}.\ is
generalized to include the strong coupling effects 
associated with in-plane interactions. The equations for the 
superconducting transition temperature $T_{c}$ are solved numerically for
several models of electron-optical phonon coupling.  
The nonmagnetic in-plane impurity
scattering suppresses $T_{c}$ in all cases considered, and  
it is possible to obtain a fair agreement with experiments for a
reasonable choice of parameters. For the anisotropic electron-phonon coupling
proposed by Song and Annett we find that the interlayer pair tunneling can 
stabilize the d$_{x^2-y^2}$-wave superconducting state with a high $T_{c}$.
Moreover, in this case there is 
a possibility of an impurity induced crossover from the d$_{x^2-y^2}$-wave state stabilized 
by the interlayer tunneling to the s-wave state at a low impurity concentration.
We also calculate the isotope effect associated with the in-plane oxygen 
optic mode 
and its dependence on the strength of 
the interlayer pair tunneling. Small positive values of the isotope 
exponent are obtained for strengths of pair tunneling that give 
high transition temperatures.
\end{abstract}
%
\narrowtext
\section{Introduction}

One of the theories for high-$T_{c}$ copper-oxide 
superconductors is the interlayer pair tunneling (ILPT) model first 
proposed by Wheatly, Hsu and Anderson \cite{Hsu} and later refined by 
Chakravarty, 
Sudb\o, Anderson and Strong \cite{CSAS,SCSA,SudboI,SudboII}. 
Recently, Chakravarty and Anderson \cite{CA} gave an indirect justification 
of this model based on a non-Fermi liquid form of the normal state 
electron propagator.  
In the ILPT-model the pairing in the individual copper oxygen layers 
is enhanced and sustained by the pair tunneling between  
the layers within the unit cell. The specific symmetry of the
component of the order parameter resulting from in-plane interactions is not 
an essential feature of the model, although in the original work \cite{CSAS}
it was assumed that this component has s-wave symmetry. In this paper 
we generalize the BCS-like form of the ILPT-model given by Chakravarty 
{\em et al}.\ to include the retardation 
(i.\ e.\ the strong coupling) effects resulting from in-plane interactions. 
This generalization is necessary in order to obtain 
a more realistic dependence of $T_{c}$ 
and other quantities characterizing the superconducting state on the 
interaction parameters \cite{Scalapino1}.
Moreover, the strong coupling form of the interlayer pair  
tunneling model is suitable for including the effect of in-plane nonmagnetic 
impurity scattering. The dependence of $T_{c}$ on impurity concentration   
 \cite{Dynes,Narlikar,Uchida,Tolpygo} is considered to be an important 
indicator of 
the symmetry of the order parameter in oxide superconductors and of the 
underlying pairing mechanism \cite{Norman,Pines}.

The self-energy equations derived in this paper are valid for any  
kind of in-plane pairing interaction within the one-boson exchange 
approximation. 
In our numerical work, however, we consider only the case when the in-plane 
pairing arises from  electron coupling to optical phonons. 
This was motivated 
by the fact that the ILPT-mechanism is novel enough that its consequences 
should be examined first when the in-plane
pairing is caused by the conventional electron-phonon interaction before more
exotic in-plane interaction models are considered. Also, 
Song and Annett \cite{Annett} recently   
derived an effective single-band Hubbard-type Hamiltonian for CuO$_{2}$ 
planes which includes the electron-phonon coupling to oxygen breathing modes.
The electron-phonon matrix element squared is proportional to 
$\sin^{2}((k_{x}-k_{x}')/2)+\sin^{2}((k_{y}-k_{y}')/2)$, where ${\bf k}$ and 
${\bf k}'$ are the electron momenta.  
With this form of coupling Song and Annett 
initially predicted that the order parameter with d$_{x^2-y^2}$-wave 
symmetry would lead to a higher transition temperature than the order parameter 
with s-wave symmetry, because in the former case the on-site Coulomb 
repulsion $U$ becomes ineffective. Subsequently, they found this prediction 
to be erroneous, which we independently confirmed during 
the course of this study. We found, however, that the d$_{x^2-y^2}$-wave 
state could be stabilized by the interlayer tunneling.  

The rest of the paper is organized as follows. In Section II we give in some  
detail the derivation of the interlayer tunneling contribution to the electron 
self-energy in the superconducting state and list the well known results for  
the self-energy parts arising from in-plane interactions. In addition, we summarize  
the $T_{c}$-equations for the case of pairing induced 
by electron-optic phonon coupling. Section III contains our numerical 
results for the transition temperature as a function of increasing disorder and 
the oxygen isotope exponent, and finally
in Section IV we give conclusions.
\section{Theory} 

\subsection{Self-energy due to interlayer pair tunneling}
In the ILPT-model it is assumed that in the superconducting state the
quasiparticle picture is approximately valid for motion within a layer, 
while the coherent motion of quasiparticles from layer to layer 
within the unit cell is blocked 
\cite{CSAS,SCSA}. The part of the Hamiltonian which describes the interlayer 
pair tunneling for two layers per unit cell \cite{CSAS,SudboII} is
\begin{equation}
H_{J}=-\sum_{\bf k}T_{J}({\bf k})[c_{{\bf k}\uparrow}^{(1)\dag}
c_{-{\bf k}\downarrow}^{(1)\dag}c_{-{\bf k}\downarrow}^{(2)}
c_{{\bf k}\uparrow}^{(2)}+H.c.]\>,
\end{equation}
where $c_{{\bf k}\uparrow}^{(i)\dag}$ is an electron creation operator for 
the state of two-dimensional momentum ${\bf k}$ and spin $\uparrow$ in the 
layer $(i)$, and 
\begin{equation}
T_{J}({\bf k})=\frac{t_{\bot}^{2}}{16t}[\cos(k_{x}a)-\cos(k_{y}a)]^{2}\>,
\end{equation}
as suggested by band structure calculations \cite{OKA}.
Here, $t_{\bot}$ characterizes the high energy single-electron coherent 
hopping from layer to layer, and is estimated to be between 0.1 eV and 
0.15 eV \cite{CSAS}. The parameter $t$ enters the tight-binding dispersion
for the electron motion within a layer 
\begin{equation}
\varepsilon_{\bf k}=-2t[\cos(k_{x}a)+\cos(k_{y}a)]-
4t^{\prime}\cos(k_{x}a)\cos(k_{y}a)-\mu\> 
\end{equation}
where $\mu$ is the chemical potential and $a$ is the lattice constant. 
To find the contribution to the anomalous electron self-energy from the 
Hamiltonian (1), it is convenient to rewrite $H_{J}$ in the Nambu formalism 
\cite{Schrieffer}. Introducing the Nambu fields 
\[
\Psi_{\bf k}^{(i)}=\left(\begin{array}{c}
		     c_{{\bf k}\uparrow}^{(i)} \\
		     c_{-{\bf k}\downarrow}^{(i)\dag}
                         \end{array} \right)\>, 
\Psi_{\bf k}^{(i)\dag}=(c_{{\bf k}\uparrow}^{(i)\dag}
\>\>\>c_{-{\bf k}\downarrow}^{(i)})
\]
the Hamiltonian (1) could be written as 
\begin{equation}
H_{J}=-\sum_{\bf k}\frac{T_{J}({\bf k})}{2}[\Psi_{\bf k}^{(1)\dag}
\hat{\tau}_{1}
\Psi_{\bf k}^{(1)}\Psi_{\bf k}^{(2)\dag}\hat{\tau}_{1}\Psi_{\bf k}^{(2)}+
				    \Psi_{\bf k}^{(1)\dag}\hat{\tau}_{2}
\Psi_{\bf k}^{(1)}\Psi_{\bf k}^{(2)\dag}\hat{\tau}_{2}\Psi_{\bf k}^{(2)}]\>,
\end{equation}
where $\hat{\tau}_{1}$ and $\hat{\tau}_{2}$ are the two off-diagonal Pauli 
matrices \cite{Schrieffer}. 
This expression looks like the sum of two two-body interaction Hamiltonians with the 
interaction line $-T_{J}({\bf k})\delta_{{\bf k},{\bf k'}}\delta_{{\bf q},0}$ 
and with the interaction vertices $\hat{\tau}_{1}$ and $\hat{\tau}_{2}$, 
respectively 
\cite{Schrieffer}. Since the ILPT-model does not consider correlations of 
the type $-\langle T_{\tau}(\Psi_{\bf k}^{(1)}(\tau)\Psi_{\bf k}^{(2)\dag}(0))
\rangle$, where $T_{\tau}$ is the Wick's time-ordering operator, we consider 
only the contribution to the electron Nambu self-energy arising from the 
Hartree-type diagrams shown in Fig.~1. The contribution of these two diagrams 
to the irreducible Nambu electron self-energy in layer (1) is 
\begin{equation}
\hat{\Sigma}_{J}^{(1)}({\bf k})=-\frac{T_{J}({\bf k})}{2}T 
\sum_{m}\left[\hat{\tau}_{1}{\rm Tr}
\{\hat{\tau}_{1}\hat{G}^{(2)}({\bf k},i\omega_{m})\}+ 
\hat{\tau}_{2}{\rm Tr}\{\hat{\tau}_{2}\hat{G}^{(2)}({\bf k},i\omega_{m})\}
\right]\>,
\end{equation}
where $T$ is the temperature in energy units, 
$\hat{G}^{(2)}({\bf k},i\omega_{m})$ is the Nambu $2\times 2$
electron Green's function for layer (2) at wave vector ${\bf k}$ and the 
fermion Matsubara frequency $i\omega_{m}$, 
${\rm Tr}\{\cdots\}$ is the trace, and the overall minus sign arises from one 
closed fermion loop. With the usual form for the {\em total}
irreducible electron Nambu self-energy in layer $i$ 
\begin{equation}
\hat{\Sigma}^{(i)}({\bf k},i\omega_{n})=
i\omega_{n}\left(1-Z^{(i)}({\bf k},i\omega_{n})\right)\hat{\tau}_{0} + 
\phi^{(i)}({\bf k},i\omega_{n})\hat{\tau}_{1} + 
\bar{\phi}^{(i)}({\bf k},i\omega_{n})\hat{\tau}_{2} + 
\chi^{(i)}({\bf k},i\omega_{n})\hat{\tau}_{3}\>,
\end{equation}
where $Z^{(i)}$ is the renormalization function, 
$\phi^{(i)}$ and $\bar{\phi}^{(i)}$
are the real and imaginary parts, respectively, of the pairing self-energy, and 
$\chi^{(i)}$ is the part of diagonal self-energy which is even in  
$i\omega_{n}$, Eq.~(5) takes the form
\begin{equation}
\hat{\Sigma}_{J}^{(1)}({\bf k})=T_{J}({\bf k})T
\sum_{m}\frac{\phi^{(2)}({\bf k},i\omega_{m})\hat{\tau}_{1} + 
\bar{\phi}^{(2)}({\bf k},i\omega_{m})\hat{\tau}_{2}}{  
(\omega_{m}Z^{(2)}({\bf k},i\omega_{m}))^{2}+(\varepsilon_{{\bf k}}^{(2)}+
\chi^{(2)}({\bf k},i\omega_{m}))^{2}+|\phi^{(2)}({\bf k},i\omega_{m})+
i\bar{\phi}^{(2)}({\bf k},i\omega_{m})|^{2}}\>.
\end{equation}
It should be noted that the interlayer pair tunneling does not lead to any 
frequency dependence in the self-energy, but contributes directly only to 
the pairing (i.~e.~off-diagonal) self-energy and, as emphasized in 
\cite{CSAS}, the resulting self-energy is local in ${\bf k}$. 
Moreover, in the 
weak coupling approximation for in-plane interactions $Z^{(2)}=1$, $\chi^{(2)}=
0$ and $\phi^{(2)}$ and $\bar{\phi}^{(2)}$ do not depend on Matsubara 
frequency, and the sum in (7) could be easily performed using contour 
integration \cite{Schrieffer}. One finds
\begin{equation}
\hat{\Sigma}_{J}^{(1)}({\bf k})=T_{J}({\bf k})
\left(\phi^{(2)}({\bf k})\hat{\tau}_{1}+\bar{\phi}^{(2)}({\bf k})\hat{\tau}_{2}
\right)\frac{1}{E_{\bf k}}\tanh\left(\frac{E_{\bf k}}{2T}\right)\>,
\end{equation}
where $E_{\bf k}=\sqrt{\left(\varepsilon_{{\bf k}}^{(2)}\right)^{2}+
|\phi^{(2)}({\bf k})+i\bar{\phi}^{(2)}({\bf k})|^{2}}$, which is the same 
result (with $\bar{\phi}^{(2)}$ gauged away) as that obtained by  Chakravarty,
Sudb\o, Anderson and Strong \cite{CSAS}.

\subsection{Self-energy due to in-plane interactions}
The precise form of the self-energy due to in-plane interactions depends on 
the model used and we will restrict ourselves to the case where the 
pairing interaction 
is due to one-boson (e.~g.~phonon or spin-fluctuation) exchange. The  
electron-phonon contribution to the self-energy is 
\cite{Schrieffer,Scalapino1,Allen}
\begin{equation}
\hat{\Sigma}_{ep}^{(i)}({\bf k},i\omega_{n})=-\frac{T}{N^{2}}
\sum_{{\bf k}',\lambda,m}
|g_{{\bf k}-{\bf k}',\lambda}|^{2}D_{\lambda}({\bf k}-{\bf k}',i\omega_{n}-
i\omega_{m})\hat{\tau}_{3}\hat{G}^{(i)}({\bf k}',i\omega_{m})\hat{\tau}_{3}\>, 
\end{equation}
where $N^{2}$ is the number of lattice sites, 
$g_{{\bf k}-{\bf k}',\lambda}$ is the electron-phonon matrix element 
for the momentum transfer ${\bf k}-{\bf k}'$ and the phonon polarization 
$\lambda$, $D_{\lambda}({\bf k}-{\bf k}',i\nu_{m})$ is the corresponding 
phonon propagator at the boson Matsubara frequency $i\nu_{m}$, and 
$\hat{G}^{(i)}$ is the electron propagator for the layer $i$. An analogous 
expression is obtained for the self-energy due to the exchange of 
antiferromagnetic spin-fluctuations, except that $|g|^{2}D$
is replaced by the spin-fluctuation propagator and the Pauli matrix 
$\hat{\tau}_{3}$ by the unit $2\times2$ matrix \cite{Kostur}.

In the case of phonon-mediated superconductivity one can include the effect 
of short-range Coulomb repulsion within an effective single band Hubbard 
model for copper-oxygen planes \cite{Annett}. The resulting contribution to 
the electron self-energy is
\begin{equation}
\hat{\Sigma}_{c}^{(i)}=-U\frac{T}{N^{2}}\sum_{{\bf k}',m}\hat{\tau}_{3}
\hat{G}_{od}^{(i)}({\bf k}',i\omega_{m})\hat{\tau}_{3}\>,
\end{equation}
where $U$ is the on-site Coulomb repulsion and $\hat{G}_{od}^{(i)}$ is the 
off-diagonal part of the electron Nambu Green's function \cite{Allen}. 

Finally we consider the effect of in-plane 
electron-impurity scattering. We will not consider all the possible effects 
of electron-impurity scattering in two-dimensional superconductors 
(e.~g.~ the enhancement of the Coulomb repulsion \cite{Hohn}), but will 
confine ourselves to the simplest treatment using either the second Born 
approximation or the t-matrix approximation.  
In the 
second Born approximation and assuming a constant electron-impurity matrix 
element $V_{i}=\langle{\bf k}|V_{N}|{\bf k}'\rangle$, where $V_{N}$ is the 
change in the crystal potential due to nonmagnetic impurity, the electron  
self-energy resulting from scattering off impurities is 
\begin{equation}
\hat{\Sigma}_{i}(i\omega_{n})=\frac{n_{i}V_{i}^{2}}{N^{2}}\sum_{{\bf k}}
\hat{\tau}_{3}\hat{G}^{(i)}({\bf k},i\omega_{n})\hat{\tau}_{3}\>,
\end{equation}
while in the t-matrix approximation the self-energy is given by
\begin{equation}
\hat{\Sigma}_{i}(i\omega_{n})=\frac{n_{i}V_{i}^{2}}{N^{2}}\sum_{{\bf k}}\hat{\tau}_{3}\hat{G}^{(i)}({\bf k},i\omega_{n})\hat{\tau}_{3}\left[\hat{\tau}_{0}-
\frac{V_{i}}{N^{2}}\sum_{{\bf k}}\hat{G}^{(i)}({\bf k},i\omega_{n})\hat{\tau}_{3}\right]^{-1}\>.
\end{equation}

\subsection{$T_{c}$-equations for the case of pairing due to electron-optic 
phonon coupling}
We consider the case where the in-plane pairing is mediated by an optic phonon
of energy $\Omega_{E}$. Near $T_{c}$ the pairing self-energies become 
infinitesimal and the self-energy equations could be linearized as described, for  
example, in  \cite{Allen,Kostur}. Assuming that the electron self-energies in 
each of the two layers are identical (e.~g.~$\phi^{(1)}({\bf k},i\omega_{n}) 
=\phi^{(2)}({\bf k},i\omega_{n})=\phi({\bf k},i\omega_{n})$) and defining 
\begin{equation}
u({\bf k},n)=\frac{\phi({\bf k},i\omega_{n})}{
\sqrt{(\omega_{n}Z({\bf k},i\omega_{n}))^{2}+(\varepsilon_{\bf k}+
\chi({\bf k},i\omega_{n}))^{2}}}\>,
\end{equation}
the $T_{c}$-equation reduces to the eigenvalue problem of a real symmetric
matrix
\begin{equation}
u({\bf k},n)=\sum_{{\bf k'},m}K({\bf k},n;{\bf k'},m)u({\bf k'},m)\>.
\end{equation}
The matrix $K$ consists of several parts associated with various interactions
\begin{equation}
K=K_{ep}+K_{J}+K_{c}+K_{i}\>.
\end{equation}
The electron-phonon contribution is 
\begin{equation}
K_{ep}({\bf k},n;{\bf k'},m)=\frac{T}{N^{2}}
\sum_{\lambda}\left|g_{{\bf k},{\bf k'},\lambda}\right|^{2}
\frac{2\Omega_{E}}{(\omega_{n}-\omega_{m})^{2}+\Omega_{E}^{2}}
S({\bf k},n)S({\bf k'},m)\>,
\end{equation}
where
\begin{equation}
S({\bf k},n)= 
\frac{1}{\sqrt{\left(\omega_{n}Z({\bf k},i\omega_{n})\right)^{2}+\left(
\varepsilon_{\bf k}+\chi({\bf k},i\omega_{n})\right)^{2}}}\>.
\end{equation}
The interlayer pair tunneling contribution is 
\begin{equation}
K_{J}({\bf k},n;{\bf k'},m)=\delta_{{\bf k},{\bf k'}}T T_{J}({\bf k'})
S({\bf k},n)S({\bf k'},m)\>,
\end{equation}
while the contribution due to on-site Coulomb repulsion is
\begin{equation}
K_{c}({\bf k},n;{\bf k'},m)=-\frac{T}{N^{2}}US({\bf k},n)S({\bf k'},m)\>.
\end{equation}
Treating the in-plane impurity scattering in the second Born approximation 
gives
\begin{equation}
K_{i}({\bf k},n;{\bf k'},m)=
\frac{n_{i}V_{i}^{2}}{N^{2}}\delta_{n,m}S({\bf k},n)S({\bf k'},m)\>,
\end{equation} 
while the t-matrix approximation gives
\begin{equation}
K_{i}({\bf k},n;{\bf k'},m)  =    
\frac{n_{i}V_{i}^{2}}{N^{2}}\delta_{n,m}S({\bf k},n)S({\bf k'},m)/D(n)\>, 
\end{equation}
where 
\begin{equation}
D(n)=\left\{\left[1+\frac{V_{i}}{N^{2}}\sum_{{\bf q}}
\left(\varepsilon_{{\bf q}}+\chi({\bf q},i\omega_{n})\right)S({\bf q},n)^{2}
\right]^{2}+
\left[\frac{V_{i}}{N^{2}}\sum_{{\bf q}}\omega_{n}Z({\bf q},i\omega_{n})
S({\bf q},n)^{2}
\right]^{2}
\right\}\>.
\end{equation}
At a given temperature the functions $Z({\bf k},i\omega_n)$, $\chi({\bf k},
i\omega_{n})$ and the chemical potential $\mu$ (see Eq.~(3)) 
are determined self-consistently by solving a set of equations 
\begin{eqnarray}
Z({\bf k},i\omega_n) & = & 1+Z_{ep}({\bf k},i\omega_n)+Z_{i}({\bf k},i\omega_n 
) \\ 
\chi({\bf k},i\omega_{n}) & = & \chi_{ep}({\bf k},i\omega_n)+
\chi_{i}({\bf k},i\omega_n)
\end{eqnarray}
together with the equation representing the particle number conservation 
\cite{Abrikosov}
\begin{eqnarray}
{\sf n} & = & \frac{1}{2}+\frac{2T}{N^{2}}\sum_{{\bf k}}\sum_{n=1}^{\infty} 
Re\{G_{1,1}({\bf k},i\omega_n)\} \nonumber \\
& = &  \frac{1}{2}-\frac{2T}{N^{2}}\sum_{{\bf k}}\sum_{n=1}^{\infty}
\left(\varepsilon_{{\bf k}}+\chi({\bf k},i\omega_{n})\right)S({\bf k},n)^{2}\>,
\end{eqnarray}
where ${\sf n}$ is the band filling factor 
and $G_{1,1}$ is the $(1,1)$-component of the electron Nambu Green's function. 
In Eqs.~(23-24) $Z_{ep}$ and $\chi_{ep}$ are given by 
\begin{eqnarray}
Z_{ep}({\bf k},i\omega_n) & = & \frac{T}{\omega_n N^{2}}\sum_{{\bf k'},m}
\sum_{\lambda}\left|g_{{\bf k},{\bf k'},\lambda}\right|^{2}
\frac{2\Omega_{E}}{(\omega_{n}-\omega_{m})^{2}+\Omega_{E}^{2}} 
\omega_mZ({\bf k},i\omega_m)S({\bf k'},i\omega_m)^{2} \\
\chi_{ep}({\bf k},i\omega_n) & = & -\frac{T}{N^{2}}\sum_{{\bf k'},m}
\frac{2\Omega_{E}}{(\omega_{n}-\omega_{m})^{2}+\Omega_{E}^{2}}
\left(\varepsilon_{{\bf k'}}+\chi({\bf k'},i\omega_{m})\right)S({\bf k'},m)^{2}
\end{eqnarray}
while $Z_{i}$ and $\chi_{i}$ are given by 
\begin{eqnarray}
Z_{i}({\bf k},i\omega_n) & = & \frac{n_{i}V_{i}^{2}}{N^{2}}\sum_{{\bf k'}}
Z({\bf k'},i\omega_n)S({\bf k'},n)^{2} \\
\chi_{i}({\bf k},i\omega_n) & = & -\frac{n_{i}V_{i}^{2}}{N^{2}}\sum_{{\bf k'}}
\left(\varepsilon_{{\bf k'}}+\chi({\bf k'},i\omega_{n})\right)S({\bf k'},n)^{2}
\end{eqnarray}
in the second Born approximation, and by
\begin{eqnarray}
Z_{i}({\bf k},i\omega_n)    & = & \frac{n_{i}V_{i}^{2}}{D(n)N^{2}}
\sum_{{\bf k'}}Z({\bf k'},i\omega_n)S({\bf k'},n)^{2}  \\
\chi_{i}({\bf k},i\omega_n) & = & -\frac{n_{i}V_{i}^{2}}{D(n)}
\Biggl[V_{i}\left(\frac{1}{N^{2}}\sum_{{\bf q}}\omega_n Z({\bf q},i\omega_n)
S({\bf q},n)^{2}\right)^{2}\Biggr. \nonumber \\
                            &   &\Biggl. \mbox{}+\frac{1}{N^{2}}
\sum_{{\bf q}}
\left(\varepsilon_{{\bf q}}+\chi({\bf q},i\omega_{n})\right)S({\bf q},n)^{2}
\left(1+\frac{V_{i}}{N^{2}}\sum_{{\bf q}}
\left(\varepsilon_{{\bf q}}+\chi({\bf q},i\omega_{n})\right)S({\bf q},n)^{2}
\right)\Biggr]
\end{eqnarray}
in the t-matrix approximation. 
The transition temperature $T_{c}$ is determined as the highest 
temperature at which the largest eigenvalue of matrix $K$ is equal to $1$ 
(see Eq.~(14)).

\section{Numerical Results}

In the numerical calculations we have taken (for definiteness) the same band 
parameters as those in the work of Chakravarty {\em et al}.; namely $t=0.25$ eV and 
$t'/t=-0.45$. The band filling factor was set at ${\sf n}=0.375$ corresponding 
to $0.75$ electrons per cell. In Fig.\ 2 we show the density of states $N(E)$ for 
these parameters obtained by adapting the tetrahedron method \cite{tetra} to a  
$400\times400$ square lattice. We assumed that the electrons couple to 
an optic phonon of energy $\Omega_{E}=62$ meV (i.\ e.\ $500\>cm^{-1}$)  
\cite{Annett}. Two models for the electron-phonon matrix element were 
considered. In the first model, which we will refer to as the isotropic model,  
the momentum-independent $\left|g_{{\bf k},{\bf k'},\lambda}\right|^{2}$ was 
assumed 
\begin{equation}
\sum_{\lambda}\left|g_{{\bf k},{\bf k'},\lambda}\right|^{2}=
	      \left|g\right|^{2}\>,
\end{equation}
and the value of $\left|g\right|^{2}$ was chosen such that the electron-phonon 
mass renormalization parameter $\lambda\approx 0.5$ .  
In the second model, which 
we will refer to as the anisotropic model, the momentum dependence of 
$\left|g_{{\bf k},{\bf k'},\lambda}\right|^{2}$ was taken to have the form 
given by Song and Annett \cite{Annett}
\begin{equation}
\sum_{\lambda}\left|g_{{\bf k},{\bf k'},\lambda}\right|^{2}=
\frac{\left|g\right|^{2}}{2}\left[
\sin^{2}(\frac{k_{x}-k_{x}'}{2})+\sin^{2}(\frac{k_{y}-k_{y}'}{2})\right]\>,
\end{equation}
with the same value of $\left|g\right|^{2}$ as in (32), so that the maximum 
in (33) is equal to $\left|g\right|^{2}$ in the 
isotropic model. Because the interlayer pair tunneling contribution to the
kernel in the $T_{c}$-equation (14) is local in ${\bf k}$, Eq.\ (19), 
the calculations had to be performed in 
${\bf k}$-space except when considering isotropic in-plane interaction with $T_{J}=0$.   
In this case it is possible to convert the ${\bf k}$-sums 
into integrals over the electron energies and use the electronic 
density of states calculated for a large ($400\times400$) lattice. The 
results for $T_{J}=0$ and isotropic in-plane interaction served as a check 
of the accuracy of the results obtained from the calculations in ${\bf k}$-space.
Due to memory size restrictions on the computers that were available to us 
(4 processor SGI R4400 and Fujitsu VPX240/10) the largest lattice size that we
could consider was $64\times64$. We found that it is absolutely critical 
to add and subtract the noninteracting form of the band filling factor ${\sf n}$ to
the expression Eq.\ (25) and to evaluate the added part as $\int dEN(E)/(\exp((E-\mu)/T)+1)$.
Otherwise, the truncation of the sum over the Matsubara frequencies in (25) and the 
finite lattice size could lead to an error in $T_{c}$ as great as $55\%$ in the 
case of isotropic in-plane interaction with $T_{J}=0$. This error in $T_{c}$ is 
largely due to error in the chemical potential $\mu$ which leads to an incorrect value for 
the density of states near $\mu$. We found that this trick of adding and subtracting 
leads to values of $T_{c}$ that are accurate to better than $5\%$ for the largest
lattice size that we could consider. The largest eigenvalue of the matrix $K$ in (14) was
obtained using the power method \cite{power} and due to the simple structure of the 
sums over the Matsubara frequencies in (25-31) there was no need to use the fast 
Fourier transform technique of Serene and Hess \cite{Serene}. The resulting code 
vectorized $93$-$97\%$ on Fujitsu VPX240/10.

\subsection{$T_{c}$ suppression by in-plane impurity scattering}

We first consider the isotropic model of electron-phonon interaction and, naturally, assume the 
s-wave symmetry of the pairing self-energy. Fig.\ 3 illustrates the suppression of $T_{c}$ by 
in-plane impurity scattering obtained within the Born approximation for four 
different values of the interlayer pair tunneling parameter $T_{J}=
t_{\bot}^{2}/t$ (Eq.\ (2)), and with the on-site Coulomb repulsion $U=0$. It should be stated from 
the outset that with the strong coupling effects (i.\ e.\ renormalization) one needs a larger 
value of $T_{J}$ to achieve a transition temperature of about 100$K$ with  
no disorder than in the BCS-like
treatment \cite{CSAS,bang} (here the electron phonon mass renormalization parameter is $\lambda=0.48$
 as deduced from the value of Z at the first Matsubara frequency and $\Omega_{E}=62$).  
In the Born 
approximation the impurity scattering is parametrized by $n_{i}V_{i}^{2}$, and we plot
$T_{c}$ as a function of this quantity. If we take $1/2\tau_{i}\equiv \pi N(E_{F})n_{i}V_{i}^{2}
$, where $N(E_{F})$ is the band electronic density of states at the Fermi level, as the measure 
of the elastic scattering rate, the range shown in Fig.\ 3 corresponds to about $110 meV$; with 
$V_{i}=t=250meV$  the maximum value of $n_{i}V_{i}^{2}$ in Fig.\ 3 is obtained for the in-plane 
impurity concentration $n_{i}=0.48$ per cell. The overall shapes 
of the curves in Fig.\ 3 are similar to the results obtained by Bang \cite{bang} in the BCS-type
of treatment using the circular Fermi surface and $T_{J}(\bf{k})\propto |\cos 2\phi|$, where 
$\phi$ gives the position of $\bf{k}$ on the Fermi surface. However, we find that $T_{c}$ is 
 suppressed at a much slower rate than what was obtained by Bang in the Born limit \cite{bang}.
At first, 
it is surprising that we get a drop in $T_{c}$ with increasing $n_{i}V_{i}^{2}$ for $T_{J}=0$. 
In this case there is no gap anisotropy which could be washed out by the impurity 
scattering leading to the suppression of $T_{c}$. Also, the structure in $N(E)$ within a range 
$\pm \Omega_{E}$ around the Fermi level does not seem to be significant enough 
(see the inset in Fig.\ 2) 
for the smearing caused by the elastic scattering rate $1/2\tau_{i}=O(\Omega_{E})$ to have 
any significant effect on $T_{c}$.  
We checked the result  for $T_{J}=0$ by converting the ${\bf k}$-sums into integrals over 
electronic energies, as discussed at the beginning of this section, and found the same result. 
Upon inspection we found that with increasing $n_{i}V_{i}^{2}$ there is a slight shift in the 
chemical potential to the region of lower density of states. Although the reduction of the 
density of states at the chemical potential is small, the exponential dependence of $T_{c}$ on 
the interaction parameters presumably leads to the observed decrease in $T_{c}$ for $T_{J}=0$. 
Note that the rate of suppression is greater for increased $T_{J}$. The reason for this trend has been 
discussed by Bang \cite{bang}.

Next we consider the effect of in-plane disorder in the t-matrix approximation. The results 
are shown in Fig.\ 4 for $T_{J}=90 meV$. 
We plot $T_{c}$ as a function of the in-plane impurity concentration 
$n_{i}$ for several values of the impurity scattering potential parameter $V_{i}$. Note that 
for $V_{i}=t=250 meV$ the t-matrix approximation (solid line) and the Born approximation (dots) 
give very similar results as one would expect in the limit of small $V_{i}$ (see Eqs.\ (11-12)). 
Increasing $V_{i}$ leads to a more rapid suppression of $T_{c}$ with increasing impurity 
concentration, and the unitary limit is reached by $V_{i}=6t$--$8t$ with $U=0$. Note, however, the 
change in curvature of $T_{c}$ versus $n_{i}$ as $V_{i}$ is increased. This trend was 
not found in the weak coupling calculation of Bang \cite{bang} in crossover from the Born limit 
to the unitary limit. However, the overall rate of suppression of the transition temperature with 
increasing disorder that we find in the unitary limit is comparable to the rate found by 
Bang \cite{bang} for $\lambda=0.5$ (with our $N(E_{F})=1.16\times 10^{-3}$ states/cell/meV/spin 
the impurity concentration $n_{i}=0.1$ cell$^{-1}$ corresponds to the impurity scattering rate 
in the unitary limit $\Gamma_{i}\equiv n_{i}/\pi/N(E_{F})=27$ meV). We also obtained step-like 
features in the $T_{c}$-curves in the unitary limit which we are not able to associate with any 
particular feature of the model and/or the numerical procedure used. The experiments 
\cite{Dynes,Narlikar,Uchida,Tolpygo} in general do not report the data on the very fine scale over 
which we observe the steps, and only in \cite{Narlikar} was there an attempt to interpret fine 
features of the observed dependence of $T_{c}$ in Y$_{1-x}$Pr$_{x}$Ba$_{2}$Cu$_{4}$O$_{8}$ on 
Pr concentration. It should be kept in mind that only the experiment of Tolpygo {\em et al}.\ 
\cite{Tolpygo} addresses specifically the in-plane  defects in YBa$_{2}$Cu$_{3}$O$_{6+x}$ at the 
fixed carrier concentration to which our model calculations apply. The most important
aspect of Fig.\ 4 is that it illustrates the profound effect of the Coulomb interaction on the 
dependence of $T_{c}$ on the concentration of in-plane impurities. For $U=8t$ (the bandwidth) 
the solid curve in Fig.\ 4 obtained for $V_{i}=t$ in the t-matrix approximation is pushed down to
the line given by squares. The decrease in $T_{c}$ for $U=8t$ is about 5K per 1\% of 
in-plane defects, similar to the value found by Monthoux and Pines \cite{Pines} for $V_{i}=t$  
in the model of spin-fluctuation-induced superconductivity and d-wave paring, and to the value 
measured by Tolpygo {\em et al}.\ \cite{Tolpygo}. Moreover, we found that for this choice 
of parameters ($\lambda=0.48$, $U=8t=2$eV) switching off the interlayer pair tunneling reduces 
the transition temperature from 84.5K (for $T_{J}=90$meV) to 1.6K for $n_{i}=0$. This illustrates  
the remarkable effect of the interlayer pair tunneling mechanism on the enhancement of $T_{c}$.

Next, we turn to the anisotropic model of the electron-phonon coupling function, Eq.\ (33). 
The results for $T_{J}=90$meV using the t-matrix approximation are shown in Fig.\ 5. 
As we have mentioned in the Introduction we were not able to obtain 
a finite transition temperature assuming d$_{x^2-y^2}$-symmetry of the pairing self-energy for 
$T_{J}=0$ down to the lowest temperature that we could consider using the ${\bf k}$-space method 
(about 20K). However, with 
$T_{J}=90$meV we obtained a transition temperature of 114K assuming d$_{x^2-y^2}$-symmetry of 
the gap for $n_{i}=0$. 
It is interesting that the s-wave case with $U=4t$ and the d$_{x^2-y^2}$-wave   
case 
(the on-site Coulomb repulsion drops out, Eq.\ (10)) give quite a similar dependence of $T_{c}$ on $n_{i}$  
for both $V_{i}=t$ and $V_{i}=10t$. The results obtained for $V_{i}=t$ are similar to the experimental 
results on YBa$_{2}$Cu$_{3}$O$_{6+x}$ with in-plane oxygen defects \cite{Tolpygo}, although we 
find that the squares in Fig.\ 4 more closely resemble the experimental data at the highest values of 
$n_{i}$ where the data seem to fall on a curve that becomes less steep as $n_{i}$ is increased.
In the unitary limit $V_{i}=10t$ the $T_{c}$-curves initially rise with increasing $n_{i}$ and then
precipitously drop. There seems to be a common threshold $n_{i}$ beyond which superconductivity 
disappears for both the s-wave pairing with either $U=4t$ or $U=8t$ and for d$_{x^2-y^2}$-wave 
pairing. We have found a similar behavior for s-wave pairing with $U=8t$, $T_{J}=90$meV and 
$V_{i}=t$ (not shown here), except that the initial rise in $T_{c}$ is much less pronounced and 
the threshold occurs at a higher value of $n_{i}$. 
 
We would like to point out that with the
electron-phonon coupling function given by Eq.\ (33) it is possible to have an impurity induced 
crossover from the d$_{x^2-y^2}$-wave state in a very pure system to 
the s-wave state at a higher impurity 
concentration, Fig.\ 6. All of the interaction parameters for the two curves in Fig.\ 6 are the same.
The only difference is that for the solid curve the d$_{x^2-y^2}$-symmetry of the pairing self-energy
is assumed (in this case the on-site Coulomb interaction drops out, Eq.\ (10)), while for the triangles 
the s-wave symmetry is assumed. At a given impurity concentration the system will go into a
state with a higher $T_{c}$ in order to lower its free-energy.

\subsection{The isotope effect associated with the in-plane oxygen optic mode}

We have examined the isotope effect associated with the optic phonon which mediates the 
in-plane interaction. In the original work of Chakravarty {\em et al}.\ it was suggested 
that the interlayer pair tunneling mechanism could explain a small isotope effect in 
high-$T_{c}$ copper oxide superconductors simply because in the interlayer tunneling model
the most important pairing process is associated with the pair tunneling. Our results 
for the isotope exponent $\alpha=-d\ln T_{c}/d \ln M$ associated with the optic mode 
at $\Omega_{E}$ are shown in Fig.\ 7. In the same figure we give the corresponding 
transition temperatures. The impurity scattering was set equal to zero and the results 
for s-wave symmetry were obtained for the on-site Coulomb repulsion $U$ equal to
zero. Note that for the isotropic model of electron-phonon interaction, Eq.\ (32), we get 
the classical result $\alpha=0.5$ for $T_{J}=0$. In the same model $T_{J}=90$meV gives 
$T_{c}=108$K and $\alpha=0.18$. Turning on the on-site Coulomb repulsion to $U=8t$ (the bandwidth) 
reduces the transition temperature to $T_{c}=84.5$K and the isotope exponent to $\alpha=0.05$ 
(not shown in Fig.\ 7)--a value approximately equal to what is found for the oxygen isotope effect  
in high-$T_{c}$ Y-Ba-Cu-O systems \cite{frank}. It should be mentioned, however, that in the site 
selective oxygen isotope experiments of Nickel {\em et al}.\ \cite{nickel}, where only the oxygen in  
copper-oxygen planes is replaced by heavier isotope, a small negative isotope effect is observed 
with the partial isotope exponent $\alpha=-0.01\pm 0.004$--close to the resolution limit \cite{frank}. 
For anisotropic electron-phonon interaction, Eq.\ (33), and assuming s-wave symmetry of the gap
we generally get lower values of $\alpha$ than in the isotropic case. For $T_{J}=90$meV and $U=0$ 
the transition temperature is 123K and $\alpha=0.04$. Turning on $U=8t$ reduces the $T_{c}$ to 
78K and increases the isotope exponent to $\alpha=0.08$. For d$_{x^2-y^2}$-symmetry of the gap we 
obtain very small positive values of $\alpha$--likely smaller than the experimental resolution 
\cite{frank}.

\section{Conclusions}

We have generalized the interlayer pair tunneling model of Anderson and coworkers to include 
the retardation effects associated with in-plane interactions. Through numerical solutions of the  
$T_{c}$-equations for a model in which electrons couple to an optic phonon at 500 cm$^{-1}$ (i.\ e.\ 
62 meV) we found, without trying to fit the experiments, that a reasonable choice for the band parameters 
($t=250$meV, $t'/t=-0.45$), band filling factor (0.75 electrons per cell), electron-phonon coupling 
($\lambda=0.48$), on-site Coulomb repulsion ($U\approx$ the bandwidth), and the interlayer pair 
tunneling strength ($t_{\bot}=0.15$eV) yields results in  surprisingly good agreement with the experiments on 
both a $T_{c}$ suppression by in-plane oxygen defects \cite{Tolpygo} and the oxygen isotope effect 
\cite{frank} in YBa$_{2}$Cu$_{3}$O$_{6+x}$. The best agreement is found for the isotropic model 
of the electron-phonon coupling function with $U=8t$ 
which leads to a $T_{c}$ suppression rate of about 5K per 
1\% of the in-plane defects (with the impurity matrix element $V_{i}=t$) 
and to the oxygen isotope exponent $\alpha=0.05$. This case also best illustrates the importance 
of the interlayer pair tunneling process in raising the transition temperature, since reducing 
$t_{\bot}$ from 0.15eV (i.\ e.\ $T_{J}=90$meV) to zero decreases the $T_{c}$ from 84.5K to 1.6K. 
We also found that for the anisotropic form of the electron-phonon coupling proposed by Song and 
Annett \cite{Annett}, Eq.\ (33), the interlayer pair tunneling can stabilize the superconducting 
state with d$_{x^2-y^2}$-symmetry at a high $T_{c}$. This stabilization occurs because 
the pair tunneling contribution 
to the pairing self-energy is local in ${\bf k}$. Morever, it is possible to have impurity 
induced crossover from the d$_{x^2-y^2}$-state in a ``perfect'' sample to the s-wave state at a higher 
concentration of defects. This is illustrated in Fig.\ 6 for $T_{J}=90$meV and $U$ equal to half the 
bandwidth, but we have also found examples of such a crossover for other values of $T_{J}$ and $U$.  

\acknowledgments
We would like to thank S.\ K.\ Tolpygo for useful discussions. 
This work was supported by the Natural Sciences and
Engineering Research Council of Canada.

\begin{figure}[htbp]
\centerplot{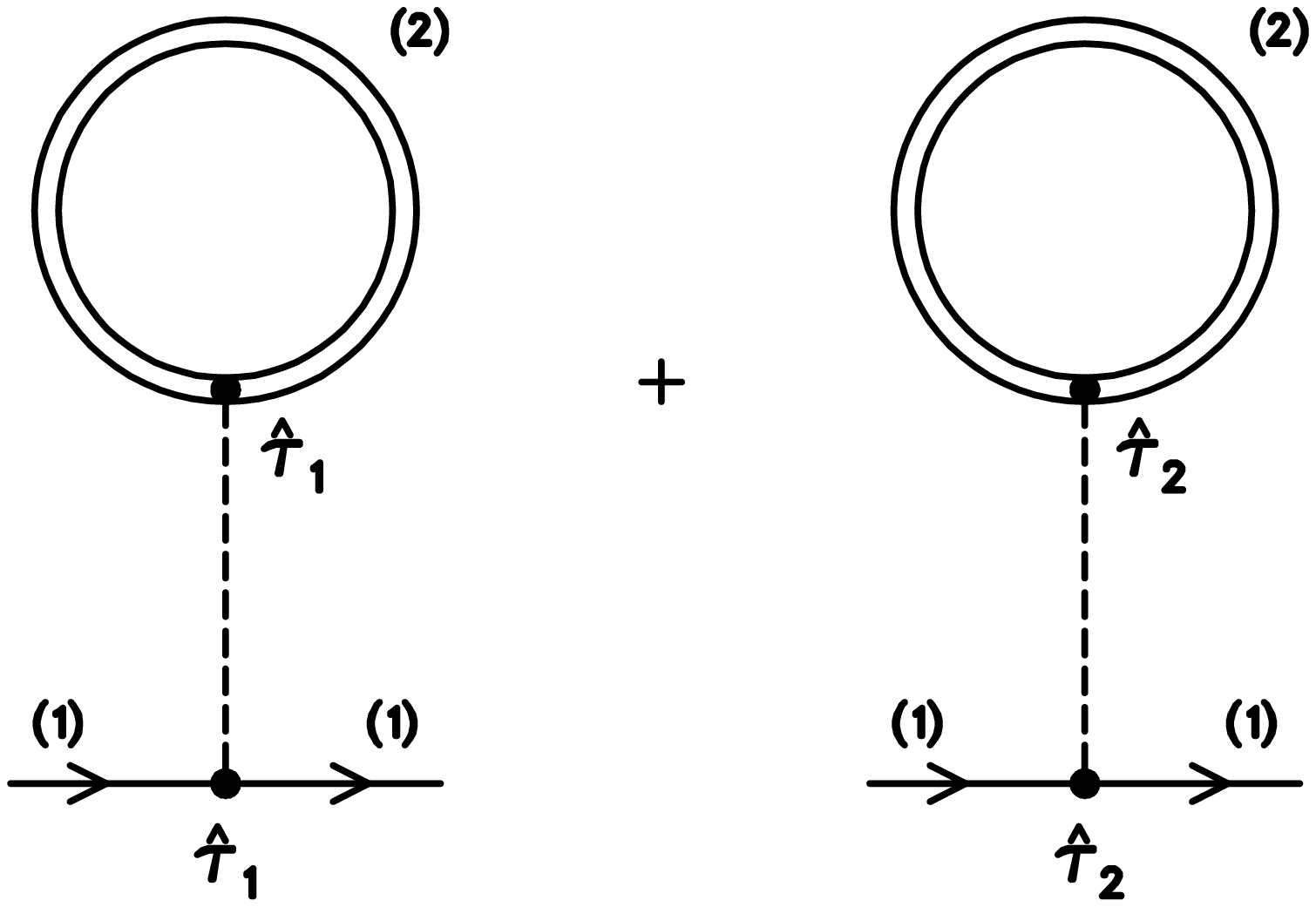}{0.5}
\caption{The interlayer pair tunneling contribution to the electron Nambu self-energy.}
\label{fig:hartree}
\end{figure}
\vspace{-0.5in}
\begin{figure}[htbp]
\centerplot{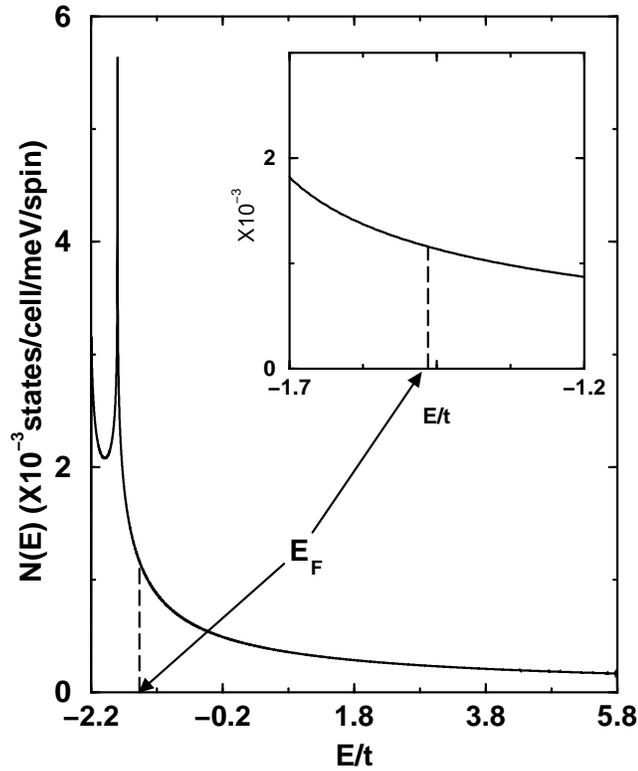}{0.55}
\caption{The electronic band density of states for the dispersion given by Eq.\
(3)
calculated for $400\times 400$ lattice using the tetrahedron method. The energy
is
measured in units of $t$ and $E_{F}$ indicates the Fermi level. The inset
shows the density of states (using the same units) in the interval
$[E_{F}-\Omega_{E},E_{F}+\Omega_{E}]$,
where $\Omega_{E}=62$meV is the energy of the optic phonon.}
\label{fig:2}
\end{figure}
\begin{figure}[htbp]
\vspace*{\fill}
\centerplot{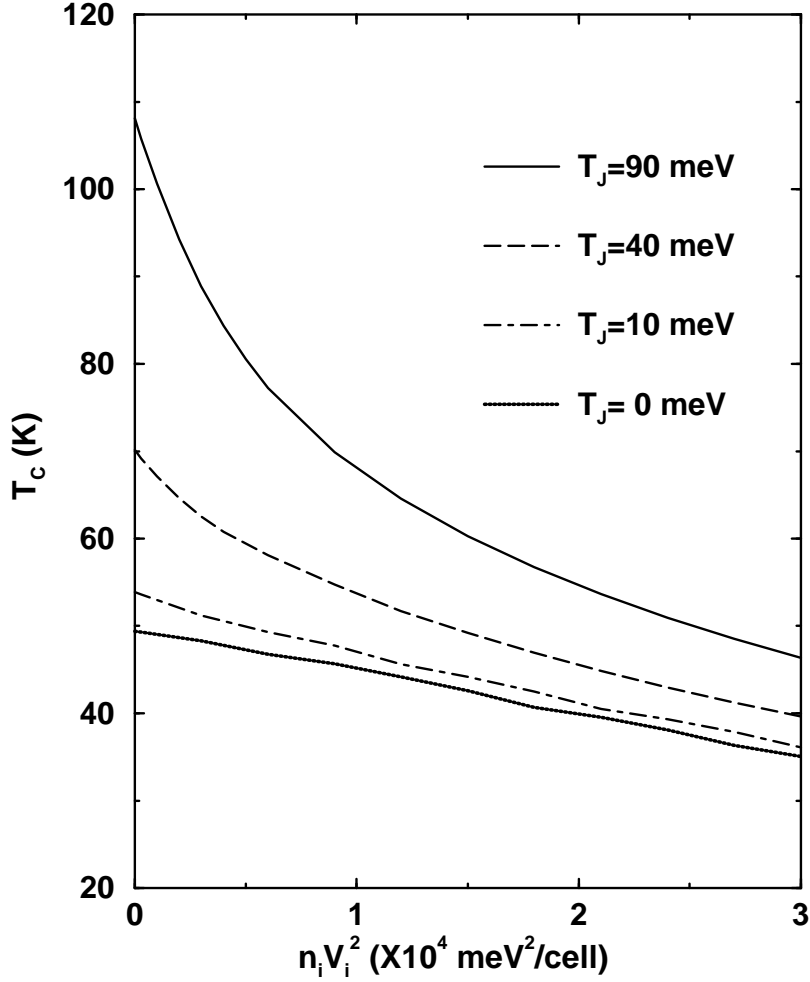}{0.6}
\caption{The dependence of $T_{c}$ on $n_{i}V_{i}^{2}$, where $n_{i}$ is the
concentration of in-plane impurities and $V_{i}$ is the electron-impurity
matrix element, calculated in the Born approximation, Eq.\ (11), for several
values of the interlayer pair tunneling strength $T_{J}=t_{\bot}^{2}/t$,
assuming the isotropic electron-phonon coupling model (32) ($\lambda=
0.48$), s-wave pairing, and the on-site Coulomb repulsion $U=0$.}
\label{fig:3}
\vspace*{\fill}
\end{figure}
\begin{figure}[htbp]
\vspace*{\fill}
\centerplot{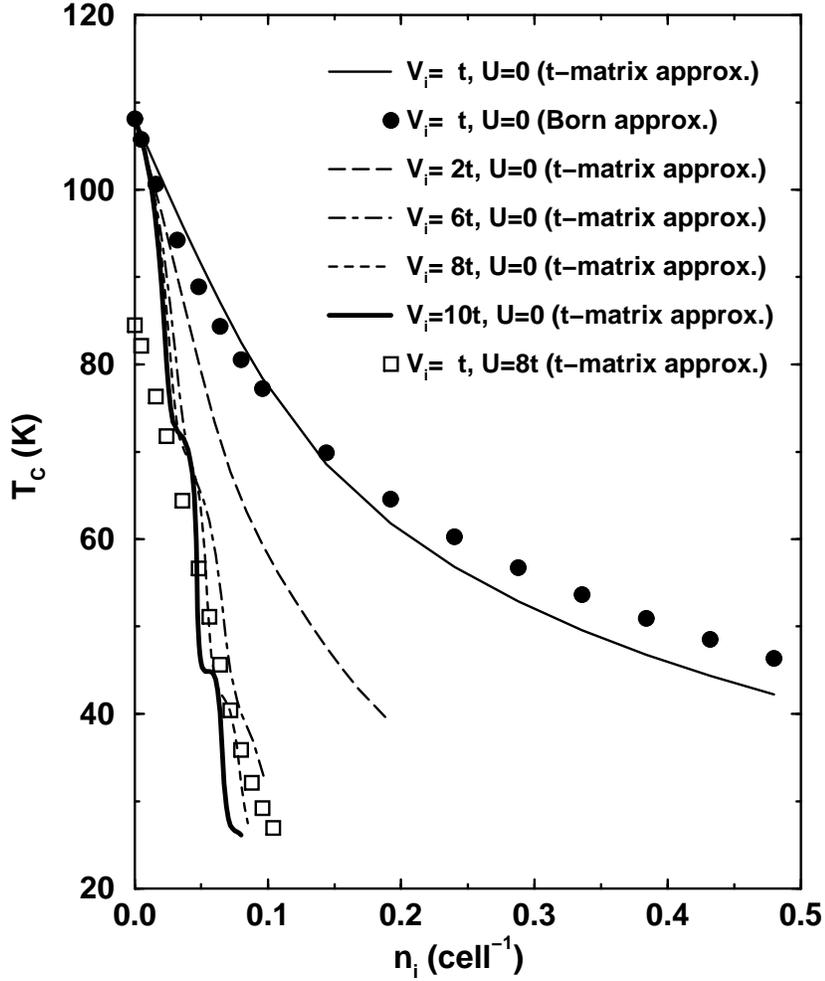}{0.6}
\caption{The dependence of $T_{c}$ on the concentration $n_{i}$ of in-plane impurities
for the isotropic electron-phonon coupling model (32) ($\lambda=0.48$) and $T_{J}=90$meV
calculated in the t-matrix approximation, Eq.\ (12), for several values of the electron-impurity
matrix element $V_{i}$, except for the results given by filled circles which were obtained in
the Born approximation (the same as the solid curve in Fig.\ 3). S-wave symmetry is
assumed and the  on-site Coulomb repulsion $U=$
except for the results given by squares
where $U$ is set equal to the bandwidth.}
\label{fig:4}
\vspace*{\fill}
\end{figure}
\begin{figure}[htbp]
\vspace*{\fill}
\centerplot{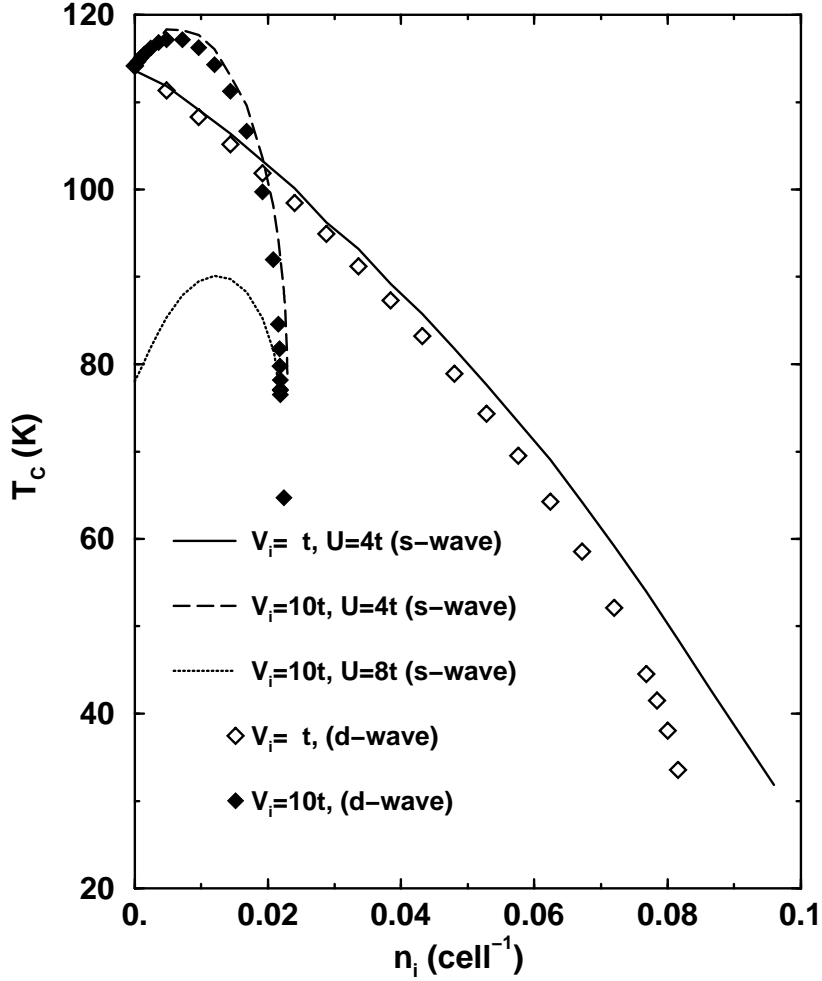}{0.6}
\caption{The dependence of $T_{c}$ on the concentration $n_{i}$ of in-plane impurities
for $T_{J}=90$meV and
for the anisotropic electron-phonon coupling model (33) (the same value of $|g|^{2}$ was
used as in Figs.\ (3) and (4)) calculated in the t-matrix approximation, Eq.\ (12),
for the electron-impurity matrix element $V_{i}$ set equal to either $t$ or
$10t$ (the unitary limit). The assumed symmetry of the gap is indicated in the
brackets and the value of the on-site Coulomb repulsion $U$
is indicated in the legend. For d$_{x^2-y^2}$-symmetry of the gap $U$ drops out, Eq.\ (10).}
\label{fig:5}
\vspace*{\fill}
\end{figure}
\begin{figure}[htbp]
\vspace*{\fill}
\centerplot{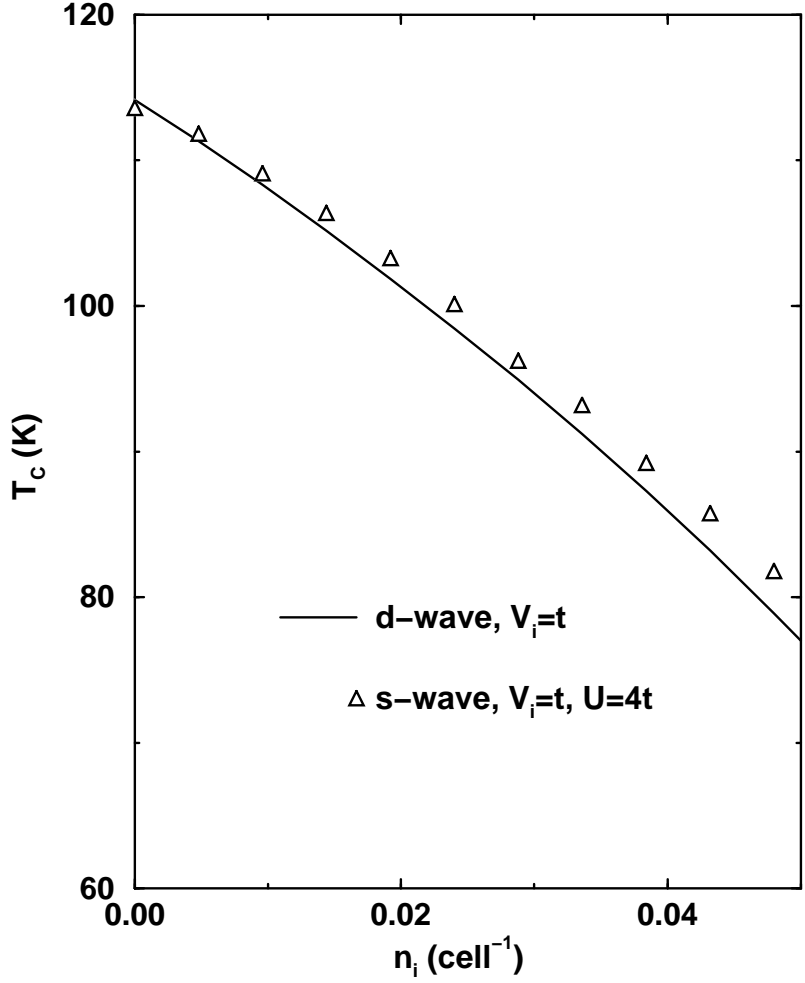}{0.6}
\caption{The dependence of $T_{c}$ on the concentration $n_{i}$ of in-plane impurities
for $T_{J}=90$meV and
for the anisotropic electron-phonon coupling model (33) (the same value of $|g|^{2}$ was
used as in Figs.\ (3), (4) and (5)) calculated in the t-matrix approximation, Eq.\ (12),
for the electron-impurity matrix element $V_{i}=t$. The results for d$_{x^2-y^2}$-symmetry
of the gap are given by the solid line and the results for s-wave symmetry are given by
triangles. Note that as $n_{i}$ increases the  state with the highest $T_{c}$ crosses over from
d$_{x^2-y^2}$-symmetry to s-wave symmetry.}
\label{fig:6}
\vspace*{\fill}
\end{figure}
\begin{figure}[htbp]
\vspace*{\fill}
\centerplot{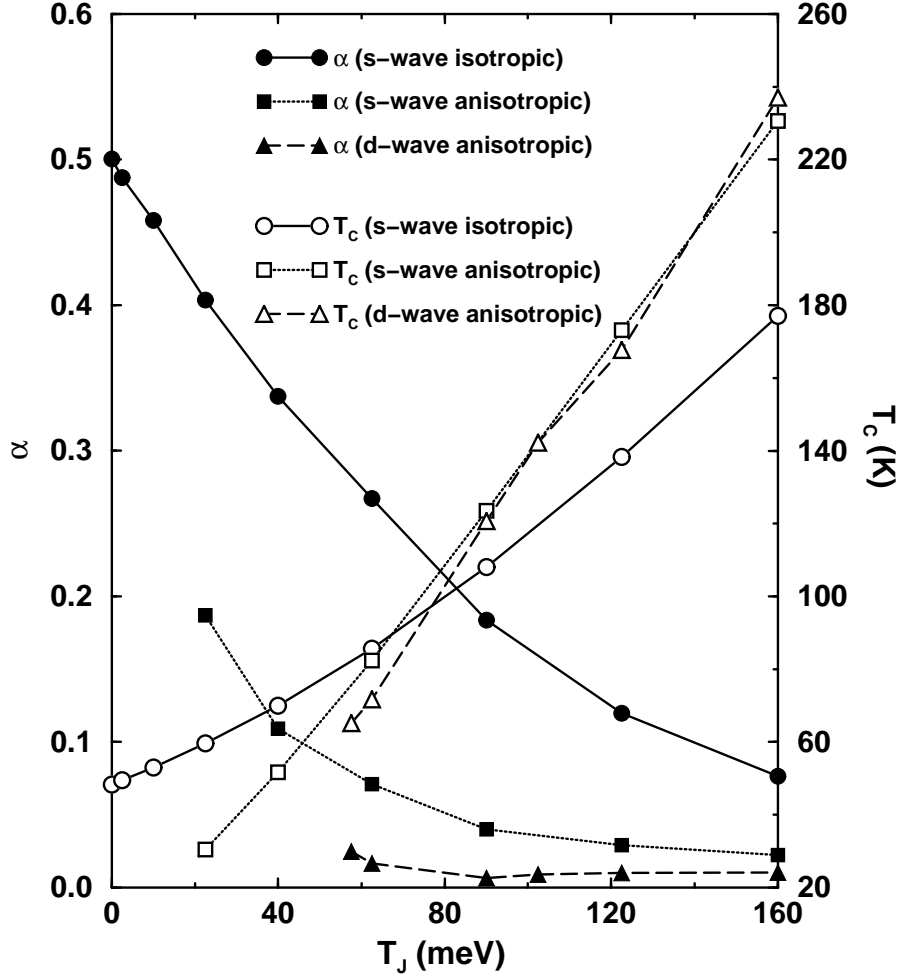}{0.6}
\caption{The isotope coefficient $\alpha=- d\ln T_{c}/d\ln M$ associated with the
oxygen optic mode at 500 cm$^{-1}$ and the corresponding $T_{c}$ as functions of the
interlayer pair tunneling strength $T_{J}=t_{\bot}^{2}/t$. In all cases the
on-site Coulomb repulsion $U$ is equal to zero. The results obtained with the
isotropic electron-phonon coupling model (32) are labeled
as (s-wave isotropic). The results obtained with the
anisotropic electron-phonon coupling model (33) are labeled as (s-wave anisotropic)
or as (d-wave anisotropic) depending on whether s-wave or d$_{x^2-y^2}$-wave
symmetry is assumed, respectively. The value of $|g|^{2}$ is the same as in
Figs.\ (3-6) leading to $\lambda=0.48$ in the isotropic case.}
\label{fig:7}
\vspace*{\fill}
\end{figure}
\end{document}